\documentclass[pdflatex,sn-mathphys-num,icol]{sn-jnl}% Math and Physical Sciences Numbered 

\usepackage{graphicx}%
\usepackage{multirow}%
\usepackage{amsmath,amssymb,amsfonts}%
\usepackage{amsthm}%
\usepackage{mathrsfs}%
\usepackage[title]{appendix}%
\usepackage{xcolor}%
\usepackage{textcomp}%
\usepackage{manyfoot}%
\usepackage{booktabs}%
\usepackage{algorithm}%
\usepackage{algorithmicx}%
\usepackage{algpseudocode}%
\usepackage{listings}%
\usepackage{siunitx}%

\usepackage{bm}

\begin{document}

\title[Article Title]{Revealing Dislocation Interactions Controlling Mechanical Properties of Metals}


\author*[1]{\fnm{Felix} \sur{Frankus}}\email{fefra@dtu.dk}

\author[2]{\fnm{Sina} \sur{Borgi}}\email{borgi@dtu.dk}

\author[2,3]{\fnm{Albert} \sur{Zelenika}}\email{albert.zelenika@kit.edu}

\author[1]{\fnm{Basit} \sur{Ali}}\email{baali@dtu.dk}

\author[4,5]{\fnm{Raquel} \sur{Rodriguez-Lamas}}\email{raquel.rodriguez-lamas@esrf.fr} 

\author[2]{\fnm{Henning Friis} \sur{Poulsen}}\email{hfpo@dtu.dk}

\author[1]{\fnm{Grethe} \sur{Winther}}\email{grwi@dtu.dk}

\affil*[1]{\orgdiv{Department of Civil and Mechanical Engineering}, \orgname{Technical University of Denmark}, \orgaddress{\street{Produktionstorvet, 425}, \city{Kongens Lyngby}, \postcode{2800}, \country{Denmark}}}

\affil[2]{\orgdiv{Department of Physics}, \orgname{Technical University of Denmark}, \orgaddress{\street{Fysikvej}, \city{Kongens Lyngby}, \postcode{2800}, \country{Denmark}}}

\affil[3]{\orgdiv{IAM-MMI Mechanics of Materials 1}, \orgname{Karlsruhe Institute of Technology}, \orgaddress{\street{Hermann-von-Helmholtz-Platz 1}, \city{Eggenstein-Leopoldshafen}, \postcode{76344}, \country{Germany}}}

\affil[4]{\orgname{European Synchrotron Radiation Facility}, \orgaddress{\street{71, avenue des Martyrs}, \city{Grenoble}, \postcode{38043}, \country{France}}}

\affil[5]{\orgname{Univ. Grenoble Alps CEA IRIG MEM NRX}, \orgaddress{\city{Grenoble}, \postcode{38000}, \country{France}}}

\abstract{During plastic deformation, metals change shape while continuously becoming stronger. The microscopic origin of these processes lies in the proliferation and movement of line defects, dislocations, and the subsequent self-organisation and pinning of dislocations on lattice imperfections, including other dislocations. The nature of these multiscale processes has remained elusive because in situ observations have not been feasible. We present 3D movies of how dislocations pile up near an obstacle, deeply within a mm-sized pure Al sample and during tensile deformation. Cross-slip is found to provide a mechanism for the dislocations to escape the pile-up, leading to pronounced intermittent behaviour. Such data support a new generation of dislocation dynamics and micro-mechanics modelling.}   

%\keywords{Dislocation dynamics, dislocation pile-up, plastic deformation}

\maketitle

\section{Introduction}\label{sec:introduction}
Metals are widely utilised in industrial applications and structural components, such as automotive bodies and bridges, due to their unique combination of mechanical strength and formability. The latter arises from the motion of dislocations, which are line defects in the crystal lattice at the atomic scale. Mechanical strength, conversely, increases when dislocation motion is impeded by interactions with obstacles, including boundaries in polycrystalline materials \cite{Hall1951, Petch1953} and other dislocations \cite{Devincre1994}.

A fundamental understanding of dislocation dynamics and interactions is therefore essential for tailoring the mechanical properties of metals. Early theoretical frameworks, developed in the mid-20th century, employed elasticity theory to derive analytical expressions that describe the stress fields and dislocation configurations resulting from dislocation interaction, e.g., \cite{Frank1950TheBoundary}. These models were later complemented by Molecular Dynamics (MD) and Discrete Dislocation Dynamics (DDD) simulations, which enabled more detailed insights into collective dislocation behaviour \cite{Kubin1992,  Arsenlis2007, Bertin2024}.

Experimentally, transmission electron microscopy (TEM) has long served as the primary tool for visualising dislocations. However, TEM is inherently limited to thin foils ($\sim$\SI{100}{\nm}), which restricts the observable volume and introduces surface effects that can alter dislocation configurations \cite{Kohnert2020, Eshelby1951XLIDislocations}. Given that dislocation lines often span micrometre-scale lengths and are attracted to free surfaces, capturing their dynamic interactions in three dimensions has remained a significant challenge.

The recent advent of synchrotron-based Dark Field X-ray Microscopy (DFXM) \cite{Simons2015,  Poulsen2021,  Zelenika2025, Lee2025} offers a transformative approach to this problem. By enabling non-destructive, three-dimensional imaging of dislocations within bulk crystals, DFXM provides unprecedented access to the spatial evolution of dislocations \cite{Jakobsen2019, Dresselhaus2021, Yildirim2023}.

In the present study, we exploit DFXM to investigate the formation, reorganisation, and sudden dissolution of a dislocation pile-up deeply embedded in a \SI{99.9999}{\percent} pure aluminium single crystal with a cross-sectional area of \SI{1}{\milli\metre\squared}, see Fig.~\ref{fig1:pile-upOverview}a. 
Dislocation migration is driven by the incremental application of macroscopic tensile deformation over 15 steps. In each step, the sample is elongated by $\Delta\varepsilon= \SI{0.02}{\percent}$ using an in situ tensile load frame. We observe that dislocation motion in the pile-up exhibits intermittency and that dislocations cross-slip to escape. The bulk nature of the sample allows for the interpretation of the data set in terms of dislocation interaction models under consistent boundary conditions.

\section{Main}\label{sec:Main}
%==========================================================================================
%   2. Main
%==========================================================================================

\subsection{Structure}
\begin{figure}
    \centering
    \includegraphics[width=1.0\linewidth]{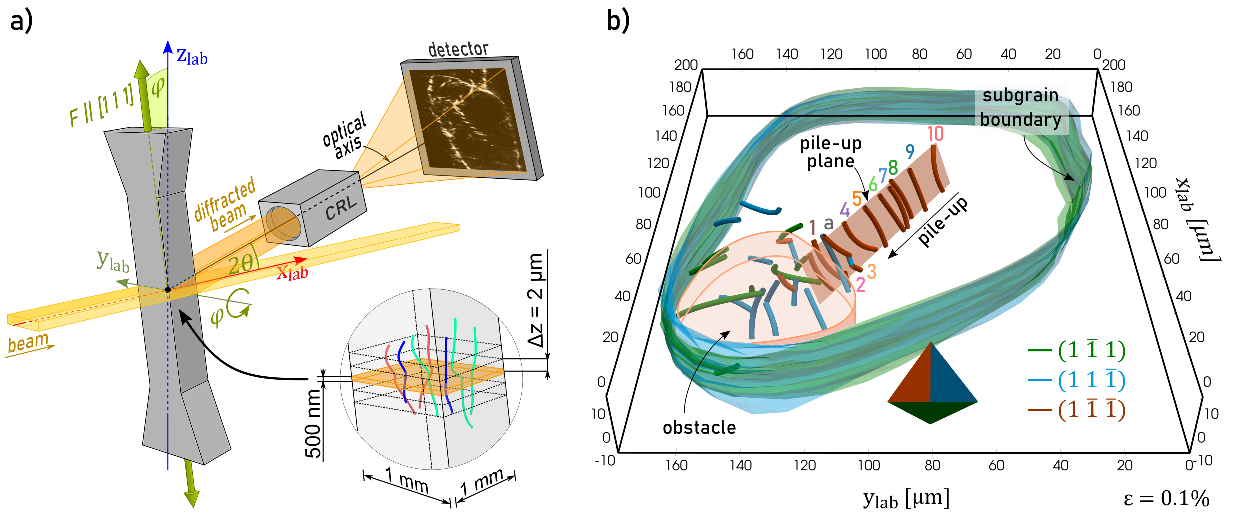}
    \caption{a) Sketch of the DFXM setup with sample orientation indicated w.r.t. the laboratory reference frame. The incident X-ray beam illuminates one layer in the sample. An X-ray microscope is inserted in the diffracted beam path, with the CRL acting as an objective, providing a magnified image of the dislocation structure within the layer. 3D information is provided by sequentially mapping layers.  b) 3D reconstruction of 10 individual dislocations piling up against an obstacle at an elongation of $\varepsilon=\SI{0.1}{\percent}$. The pile-up forms a coplanar arrangement with respect to the common slip plane $(1\bar{1}\bar{1})$. The dislocations are colour-coded according to their slip plane normal.}
    \label{fig1:pile-upOverview}
\end{figure}

The studied dislocation ensemble -- the pile-up -- is located within a subgrain (domain) at the centre of the sample's cross-section, see Fig.~\ref{fig1:pile-upOverview}b. The two tessellated representations of the boundary of this sub-grain,  green at the first and blue at the last of the 15 load increments, indicate that the shape of the boundary is invariant. 
This low-energy boundary does not impose long-range stresses on the cell's interior and shields it from stresses caused by dislocation structures outside the subgrain.

%==========================================================================================
%   2. b) Individual Dislocations
%==========================================================================================

%Contrary to the global invariant morphology of the boundary, 
The evolution of the ensemble of interacting dislocations inside this subgrain is presented in Supplementary Video~\ref{figSupp:9-Movie}. A snapshot of this video is shown for the fifth step $\varepsilon = \SI{0.1}{\percent}$ in Fig~\ref{fig1:pile-upOverview}b. The 10 dislocations numbered in the figure are likely to have formed sequentially, by the same source to the upper right of the figure. Initially, these appear on closely spaced and parallel slip-planes, where they, on average, slip (travel) in the direction of the arrow. At the displayed step, these defects form a metastable structure, a \textit{pile-up}, in which the further migration of the dislocations is impeded by an obstacle that serves as a barrier.
This pile-up represents a direct visualisation of how the dislocation mobility is reduced, which is the microscopic origin of work-hardening, the fact that metals become harder when deformed. The dislocations repel each other, leading, in classical theory, to a characteristic gradual decrease in the inter-dislocation spacing towards the obstacle and to an increasing stress in the direction of the slip. Uniquely with DFXM, we can observe such structural evolution under bulk conditions.

\subsection{Dynamics}
\begin{figure}
    \centering
    \includegraphics[width=1.0\linewidth]{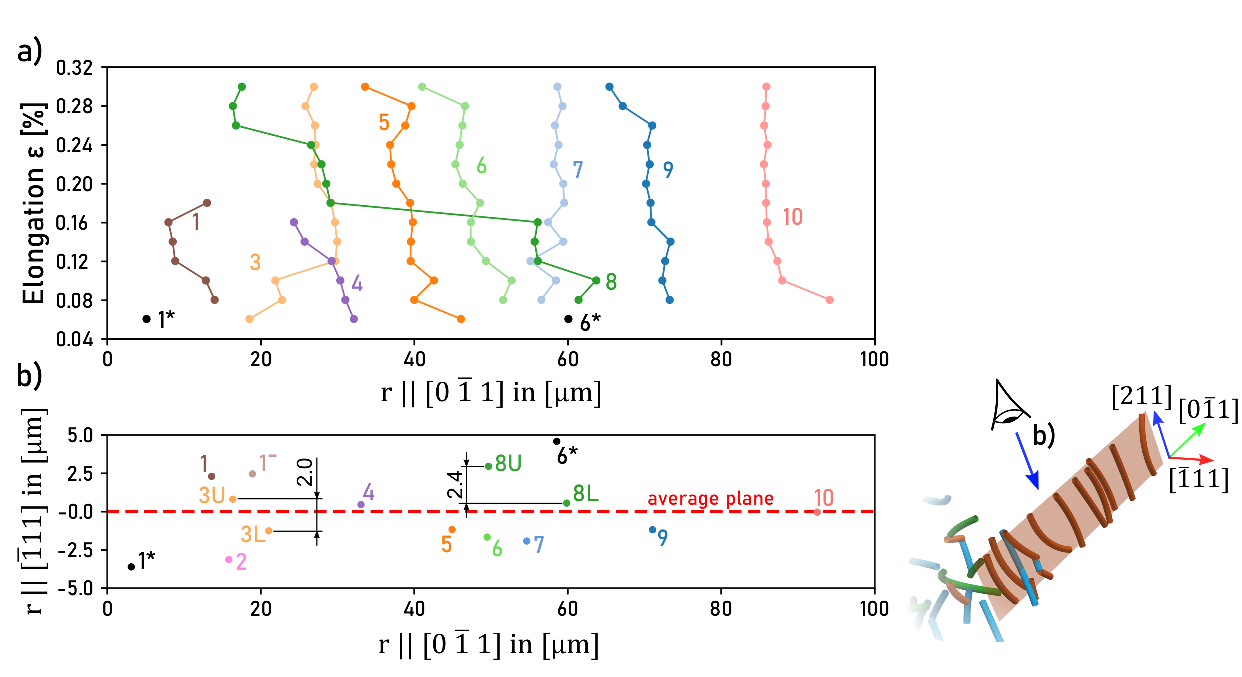}
    \caption{Projections of dislocation line positions. a) Positions at $z=\SI{0}{\micro\metre}$ for each dislocation as a function of steps (dots and lines). b) Normal distance of the dislocation lines to the average pile-up plane. The plotted positions along the pile-up trace $r\parallel[01\bar{1}]$ represent the centre of mass of the dislocation lines at the step where they first appear and thus do not correspond to a specific time step. Dislocations 3 and 8, showing cross-slip segments, are plotted twice with indexes L and U referring to the lower and upper segments, respectively. The view direction of subplot b) is indicated in the sketch in the lower right.}
    \label{fig:2-StructuralEvolution}
\end{figure}

The in situ deformation of the sample in small steps enables tracking the evolution of the pile-up configuration, as detailed in Supplementary Section~\ref{sec:appendix-evolution} and Fig.~\ref{figSupp:5-evolution}.
Overall, the accumulated migration of the dislocations over the load sequence describes a gradual motion towards the obstacle, as shown in Supplementary Fig.~\ref{figSupp:4-structuralMetrics}b. Here, the total area swept by the dislocations is seen to be an almost linear function of the applied macroscopic elongation. At the individual dislocation level, however, the response is much more stochastic in nature.
Shown in Fig.~\ref{fig:2-StructuralEvolution}a is the centre-of-mass position of the dislocations as a function of sample elongation. Notably, on some occasions, dislocations reverse the direction.
(This is also evident from 3D representations, see Supplementary Fig.~\ref{figSupp:6-positionOverlay}). Even when paths appear consistent and uniform, e.g. dislocation $5$ in Supplementary Fig.~\ref{figSupp:6-positionOverlay}, their motion is interrupted and erratic, suggesting the presence of threshold forces, a characteristic we shall term \textit{intermittency}. This challenges the conventional paradigm of dislocation dynamics, which assumes dislocation motion follows a defined functional relationship with the applied driving force.
The data also reveal that, with a certain likelihood, dislocations escape from the common slip plane by cross-slip, in which parts of the dislocation start expanding into a different plane. Despite this mode being limited to dislocations that are nearly parallel to their Burgers vector, this mechanism appears to play an essential role in dislocation migration in pure aluminium. The cross-slip distance is approximately \SI{2}{\micro\metre} for both dislocations 3 and 8, and shows a similar magnitude as the deviation of all dislocations relative to the pile-up’s average plane, cf. Fig.~\ref{fig:2-StructuralEvolution}b. Assuming the dislocations were emitted from the same source, this distance appears to be an inherent parameter of the system.
(The addition of new dislocations is not resolved by our measurements, since such events occur during elongation between single measurements.)\par

\subsection{Cross Slip Event}
\begin{figure}[h]
    \centering
    \includegraphics[width=1.0\linewidth]{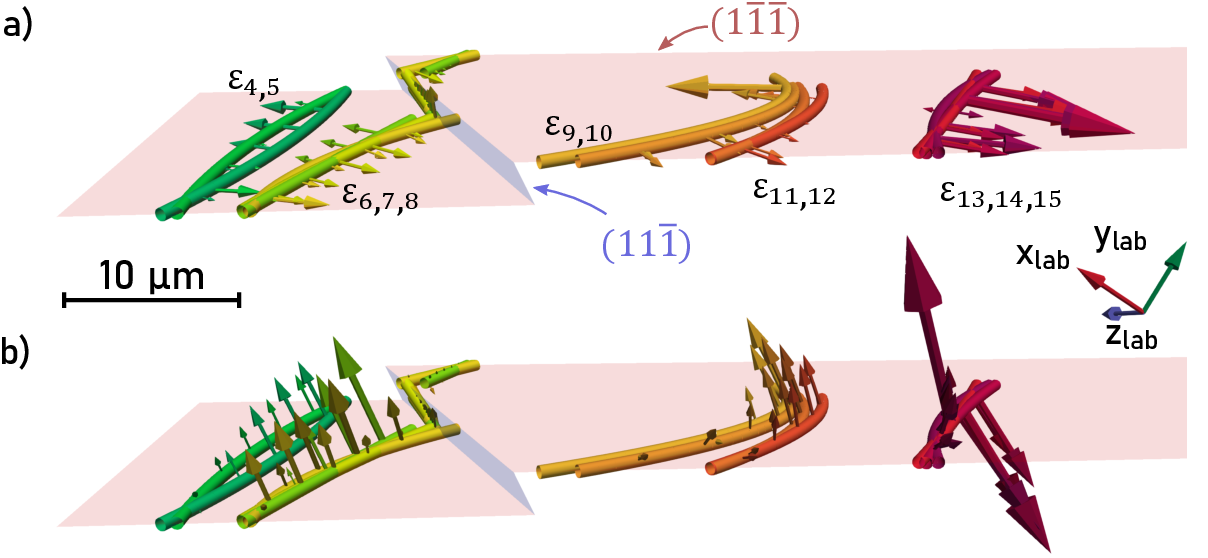}
    \caption{Step resolved reconstruction of the double cross-slip event of dislocation \textit{8}. The path of the dislocation is marked by steps 4 to 15. Between steps 5 and 6 parts of the dislocation migrate to a parallel slip plane of type $\left(1\bar{1}\bar{1}\right)$ (red) via a segment on the cross-slip plane $\left(11\bar{1}\right)$ (blue), resulting in a larger normal distance to the pile-up. The force acting on the dislocation is marked by arrows. a)  Force component within the red slip planes. b) Force component on the cross-slip plane of type $(11\bar{1})$.}
    \label{fig:3-CrossSlip}
\end{figure}

These events are exemplified by dislocation line 8 -- green in Fig. \ref{fig:2-StructuralEvolution} -- which exhibits a double cross-slip through which it leaves its slip plane, travels a short distance in another slip plane and returns to a plane parallel to its original slip plane. This enables it to overtake the neighbours in the pile-up configuration. The trajectory of this cross-slip event is shown in 3D in Fig.~\ref{fig:3-CrossSlip}.
The migration of the dislocation line involves sporadic halts for multiple steps (cf. $\varepsilon_4\rightarrow\varepsilon_5$, $\varepsilon_{6}\rightarrow\varepsilon_{8}$, $\varepsilon_{9}\rightarrow\varepsilon_{12}$, and $\varepsilon_{13}\rightarrow\varepsilon_{15}$), each followed by significant glide events over distances of up to $\SI{10}{\micro\metre}$ within a single step (cf. $\varepsilon_5\rightarrow\varepsilon_6$, $\varepsilon_8\rightarrow\varepsilon_9$, and $\varepsilon_{12}\rightarrow\varepsilon_{13}$).\par
The intermittent trajectory of the dislocation line is divided into three states. In the beginning, in steps $\varepsilon_4$ and $\varepsilon_5$, the dislocation line resides on a slip plane close to the average pile-up plane (c.f red plane in Fig.~\ref{fig1:pile-upOverview}b, its trace in Fig.~\ref{fig:2-StructuralEvolution}b, and position of lower segment 8L). As shown for step $\varepsilon_5$ in Supplementary Fig.~\ref{figSupp:5-evolution}, the dislocation line is located between dislocations 7 and 9 in the pile-up. \par
In the subsequent steps $\varepsilon_6$ to $\varepsilon_8$, the dislocation line passes by dislocation 7 on its original slip plane while the upper third of the dislocation line double cross-slips onto a parallel slip plane of type $\left(1\bar{1}\bar{1}\right)$ with a normal distance of $\|d\|$=\SI{2.4}{\micro\metre}. The cross-slip occurs through a short line segment on the slip plane $\left(11\bar{1}\right)$, indicated in blue in Fig.~\ref{fig:3-CrossSlip}. 
During the remaining steps $\varepsilon_9$ to $\varepsilon_{15}$, the dislocation line entirely resides on the newly populated slip-plane parallel to the pile-up. Here, only the lower two-thirds are visible in the data. 

\subsection{Mechanical Modelling}\label{sec:mechanicallModelling}
This data set presents an unprecedented opportunity to compare well-established dislocation interaction theory and DDD models with experimental data. To facilitate this, a mechanical model for the driving forces acting on reconstructed dislocations in each step is implemented. The force calculations are based on a non-singular solution for stress fields of finite straight dislocation segments \cite{Cai2006}, see Section~\ref{sec:force-calculations}. 
To account for long-range stresses acting on the entire sub-grain structure, we fit an effective applied load to obtain qualitative agreement between the directions of the calculated force and dislocation motion. The result is \SI{0.8}{\mega\pascal} -- well below the material’s yield point of \SI{12}{\mega\pascal}.\par

In Fig.~\ref{fig:3-CrossSlip}, the calculated forces acting on the cross-slipping dislocation \textit{8} are shown for each step, highlighting the correlation with the non-linear migration trajectory. 
The observation that the dislocation line remains stationary for certain steps aligns with changes in the force vector directions (cf. $\varepsilon_4$, $\varepsilon_6$, $\varepsilon_9$), indicating significant local variations in the stress field due to the motion of neighbouring dislocations in the pile-up. \par

The out-of-plane motion of the dislocation line during the step $\varepsilon_5\rightarrow\varepsilon_6$ aligns with the calculated force-components on the cross-slip plane $(11\bar{1})$ presented in sub-figure b). In steps $\varepsilon_6$ to $\varepsilon_8$, the lower segment of the dislocation partially experiences more than double the force along the cross-slip plane (subfigure b) compared to within its habit plane (subfigure a). The repelling out-of-plane force originates from the dislocation line's relative normal displacement of approximately \SI{2}{\micro\metre} with respect to its direct neighbours \textit{7} and \textit{9} (cf. 8L in Fig.~\ref{fig:2-StructuralEvolution}b). This dislocation-dislocation interaction force is significantly diminished for the cross-slipped upper segment (8U, same Fig.) due to its larger radial distance to its neighbours. Combined with the applied load acting towards the obstacle, these repulsive forces gradually drag the dislocation onto the energetically more favourable distant parallel $(1\bar{1}\bar{1})$ plane. Both force projections are drastically increasing as the dislocation line approaches the vicinity of the obstacle in the last steps, from $\varepsilon_{13}$ to $\varepsilon_{15}$. 

The calculated forces acting on \emph{all} dislocations are shown in Supplementary Fig.~\ref{figSupp:8-mechanicalModelling} for three selected steps. By overlaying the forces with the dislocation positions in the consecutive step, their direction and magnitude can be compared to the dislocation migration as further discussed in Supplementary Section~\ref{sec:app-forces}.

\section{Conclusions}
The use of Dark-field X-ray Imaging has, for the first time, enabled the tracking and quantification of dislocation motion and dynamics during plastic deformation under true bulk conditions. The observed evolution of the dislocation structure highlights the highly complex interactions of defects at the microscopic scale. Even in this model material with a  seemingly simple configuration, the pile-up, stochastic effects contribute significantly to the overall dynamics, giving rise to the observed characteristics: \textit{intermittency} and \textit{cross-slip}.\par 
In outlook, we note that a different DFXM modality enables mapping of the entire deformation gradient tensor field \cite{Henningsson2025, Detlefs2025}. From such data the driving forces on the imaged dislocations can be directly probed without the predicament of unknown boundary conditions arising from features outside the imaged volume. By correlating driving forces and dislocation motion, crucial yet hardly accessible parameters, such as mobility and the currently determined cross-slip distance of \num{2}-\SI{3}{\micro\metre}, 
can be directly determined and fed into DDD models.
Likewise, new X-ray optics such as Multilayer Laue Lenses \cite{Kutsal2019}, providing higher spatial resolution, suggest that applications involving much higher dislocation densities are within reach. In addition to validating and refining dislocation interaction models, such a dataset can serve as initial input for DDD simulations at higher degrees of deformation, addressing the limited deformation levels that can be effectively covered.

\section{Methods}\label{sec:methods}
\subsection{Synchrotron Experiment}
The Dark-Field X-ray Microscopy data were collected at the Hard X-Ray Microscopy Beamline ID06-HXM at the European Synchrotron Radiation Facility (ESRF), \cite{Kutsal2019}. 

\subsubsection{Sample and in situ loading equipment}
The tensile sample is an aluminium single crystal of \SI{99.9999}{\percent} purity with a parallel length of \SI{10}{\milli\metre} and a square cross-section of \SI{1}{\milli\metre\squared}. The sample had been annealed at \SI{540}{\celsius} for \SI{10}{\hour} to minimise the dislocation density as much as possible. It is mounted in a customised miniature tensile load frame allowing for in situ deformation. More details on sample orientation and the load frame are provided in Supplementary Section~\ref{sec:appendix-loadFrame}. 

\subsubsection{Diffraction Setup}
The scans are collected using a monochromatic condensed line beam with a vertical height of $\sim$\SI{500}{\nano\metre} at an X-ray energy of \SI{17.0}{\kilo\eV}. A sketch of the microscope setup is presented in Fig.~\ref{fig1:pile-upOverview}a). The probed diffraction vector $\bm{Q}$ is parallel to the sample's tensile axis with a crystallographic direction of $[111]$. The side surface of the sample with a crystallographic direction of $[\bar{1}01]$ is aligned with the y-axis of the laboratory reference frame $y_\mathrm{lab}$. The entry plane of the CRL and the detector (image plane) are mounted at distances of \SI{266}{\milli\metre} and \SI{5000}{\milli\metre} from the sample, respectively. The images are acquired with a $2160\times2560$ pixel PCO.edge sCMOS camera at an exposure time of \SI{1}{\second}. For the given energy, the diffraction angle for the $(111)$ plane is $2\theta=\SI{17.976}{\degree}$, resulting in an effective pixel size in the laboratory reference frame of $\num{57}\times\SI{17}{\nano\metre}$.

\subsubsection{Scan Procedure}
After preloading the sample until the first motion of dislocations is observed, i.e. the yield point, incremental loading in steps of $\Delta\varepsilon=\SI{0.02}{\percent}$ is performed. \par
The measurement procedure at each load step consists of layer-wise rocking scans. The rocking scans are conducted through the incremental counter-clockwise rotation of the sample stage in steps of $\Delta\varphi=\SI{0.035}{\milli\radian}$ around the laboratory y-axis $y_{\mathrm{lab}}$ over a total angular range of \SI{0.35}{\milli\radian}. As the sample is rocked, different lattice orientations are probed. By rotating the sample by a sufficiently large amount (here: $\sim\SI{15}{\milli\radian}$), contrast between the deformed lattice around dislocations and the bulk lattice is established - a mode termed weak beam. A total of 11 layers with an inter-layer spacing of \SI{2}{\micro\metre} is collected for 20 elongation steps.
\subsubsection{Data Analysis}
The collected images are converted and stacked into intensity arrays representing the scanned volume. Following primary background subtraction and manual annotation of the dislocation cores through layer-wise Gaussian blob detection, the dislocation lines are fitted as described in Supplementary Section~\ref{sec:appendix-DislocationReconstruction}.

\subsection{Force Calculations}\label{sec:force-calculations}
The dislocation interaction forces are calculated using the non-singular expressions for the stress field associated with finite straight dislocation segments presented by Cai et al. \cite{Cai2006}. For the calculations, the obtained fits of the dislocation lines (cf. Supplementary Section~\ref{sec:appendix-DislocationReconstruction}) are discretised in linear segments of length \num{0.6} to \SI{1.0}{\micro\metre} ($2100$ to $3500 \lvert\bm{b}\rvert$). A Burgers vector $\bm{b}\parallel[101]$, the only common direction to the two observed slip planes $(1\bar{1}\bar{1})$ and $(11\bar{1})$, is assumed for all segments. The assumption is justified by the similarity of the profiles on the detector images of all dislocations in the pile-up. 

To minimise the error in stress field calculations at the boundaries of the imaged volume, which is caused by the clipping of dislocation lines, the dislocation configurations are tangentially extended by approximately \SI{5}{\micro\metre}. An extended elaboration on the calculations is given in Supplementary Section \ref{sec:appendix-force}.
\section{Funding}
F.F., S.B., A.Z., G.W., and H.F.P. acknowledge funding from ERC Advanced Grant number 885022, and from the ESS lighthouse on hard materials in 3D, SOLID, funded by the Danish Agency for Science and Higher Education, Denmark, grant number 8144-00002B. H.F.P. acknowledges support from the Villum Foundation through Villum Investigator Grant number 73771. We acknowledge ESRF for the provision of synchrotron radiation facilities under proposal number MA-4442 on beamline ID06-HXM.

\section{Contributions}
F.F., S.B., A.Z., B.A., R.R.L., H.F.P., and G.W. performed the experiments. F.F. built the load frame, conducted data analysis, and developed the dislocation fitting algorithm. H.F.P. and G.W. designed the experiment. F.F., G.W., and H.F.P. interpreted the data and wrote the article, and all authors contributed and commented on the text.

\section{Competing Interests}
The authors declare no competing interests.

\clearpage

\nocite{*}
\bibliography{sn-bibliography}% common bib file
\clearpage
\begin{appendices}
\section{Supplementary Information to Methods}
\subsection{In Situ Load Frame and Sample Geometry}\label{sec:appendix-loadFrame}
\begin{figure}[h]
    \centering
    \includegraphics[width=1.0\linewidth]{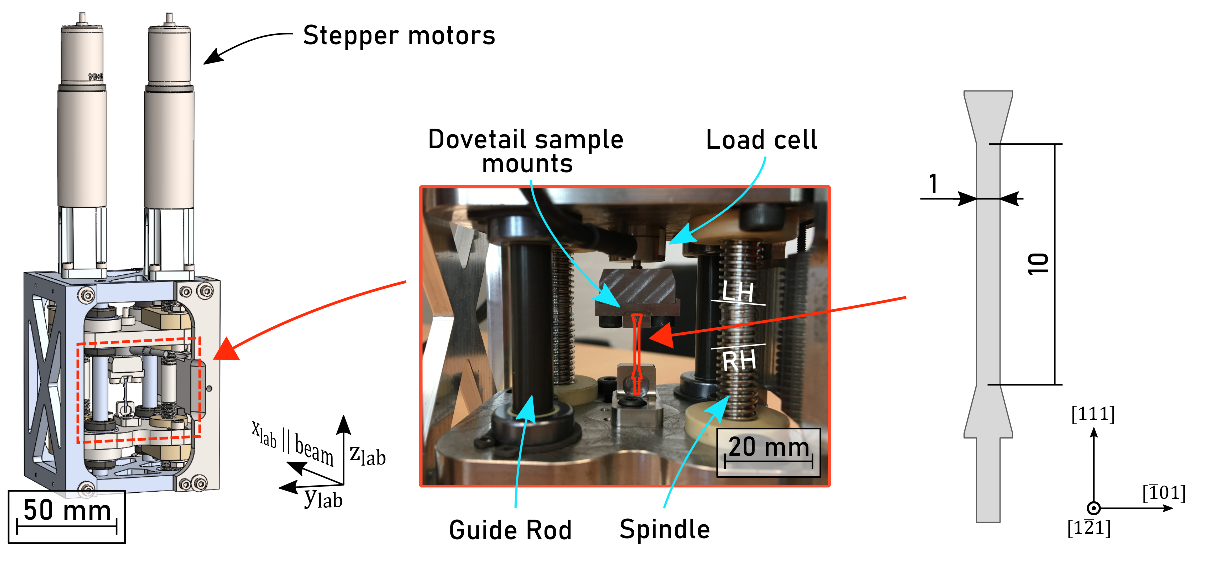}
    \caption{Sketch of in situ loading frame and aluminium single crystal tensile specimen with crystallographic orientations.}
    \label{figSupp:1-loadFrame}
\end{figure}

The load frame is a remotely operated, miniature tensile machine mounted on the goniometer stack. It features two synchronously moving traverses, driven by spindles with opposed threading. This configuration maintains a quasi-steady central section during sample elongation, minimising displacement of the gauge area. The deformation process is controlled through geared stepper motors with an effective step size of $\Delta l = \SI{0.211}{\micro\metre}$ giving rise to a minimum elongation of \SI{0.002}{\percent} for the sample's parallel length of \SI{10}{\milli\metre}. The reaction force is directly read out at the upper traverse through a load cell with a nominal load of \SI{500}{\N}. Elongations stated in this work are evaluated from the traverse displacement. An ex-situ force-elongation curve based on elongations determined using Digital Image Correlation (DIC) of an identical single crystal is shown in Fig.~\ref{figSupp:7-stress_strain_curve}. The tensile specimen is mounted through dovetail-shaped ends, providing form fit, and additionally glued. A rendering of the load frame, together with a close-up photo of the traverses, and a sketch of the sample are shown in Fig.~\ref{figSupp:1-loadFrame}.

\subsection{Data Treatment}\label{sec:appendix-dataAnalysis}
\begin{figure}[h]
    \centering
    \includegraphics[width=1.0\linewidth]{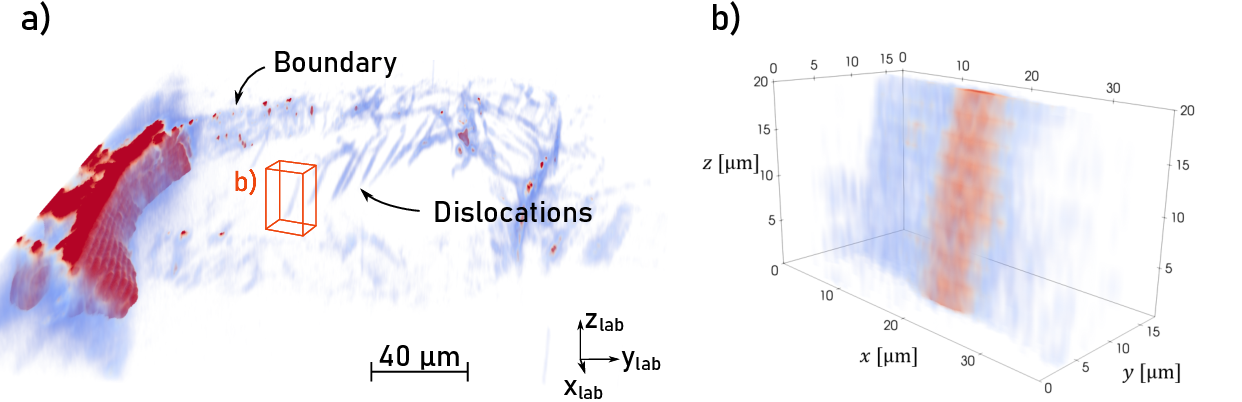}
    \caption{Intensity field in laboratory reference frame for step $\varepsilon_\mathrm{5}$ at $\Delta\phi=\SI{0.14}{\milli\radian}$. \textit{a)} full scanned volume showing dislocation boundaries and individual dislocations. \textit{b)} zoomed-in intensity field around a single dislocation $10$.}
    \label{figSupp:2-intensityField}
\end{figure}

The collected data set is a discrete 5-dimensional field of intensities $I\left(x,y,z, \phi, \varepsilon\right)$.
The dislocation reconstruction from the intensity fields is conducted through a manual fitting program implemented in Python. The reconstruction is based on characteristic, spatially extended, diffraction signals arising from the strained lattice around the dislocation core. These diffraction signals become visible as the sample is rotated away from its nominal diffraction vector, i.e., only contributions of the diffraction signal arising from the locally strained lattice reach the detector. For a given diffraction geometry, the shape of that signal is dependent on the orientation of the Burgers vector and the line direction \cite{Borgi2024}. The stacking of these signals in the volumetric scans of this dataset results in tube-shaped regions of high intensity surrounding the dislocation core. A volumetric rendering of the intensity field, along with a zoomed-in view of an individual dislocation, is shown in Fig.~\ref{figSupp:2-intensityField}. \par 
\subsection{Dislocation Reconstruction}\label{sec:appendix-DislocationReconstruction}
To recover the dislocation configuration from the intensity fields, the positions of the dislocations in each layer of the scanned volume are approximated as the centres of their characteristic diffraction signatures. The authors acknowledge that this is a strong approximation of the dislocation core position, as the anisotropic character of the deformation gradient field of the dislocation and the finite beam height result in a more complex characteristic of the imaged signature. Yet, due to the overall elliptic shape of the signatures in the image plane, the centre is determined through Gaussian blob detection. The identified points $\bm{p}_{i,k}$, which describe the position of the dislocation line $d_j$ in each layer $l_i$ and step $\varepsilon_k$, are then used for further parameterisation of the dislocation line. It is assumed that the dislocation line resides on one of the four possible $\{111\} $ planes.\par
Following this assumption, the position of the dislocation's slip plane can be determined as the centre of mass $\bm{c}_j$ of all points $\bm{p}_{i,k}$:
\begin{equation}
\label{eq:centreofmass}
\bm{c}_j=\frac{1}{N}\sum^{N}_{i=1}\bm{p}_{i,k}
\end{equation}
The slip plane normal $\hat{\bm{n}}$ is identified by constructing a set of candidate planes $S_c$, each with a normal vector $\hat{\bm{n}}_c$ aligned with one of the four $\{111\}$ variants, and evaluating the residuals relative to the observed points.
\begin{equation}
r_c=\sum^{N}_{i=1}\left|\hat{\bm{n}}_c\cdot\bm{p}_{i,k}-\left|\bm{c}_j\right|\right|\text{, with }\hat{\bm{n}}_c\in\{111\}.
\end{equation}
To further characterise the dislocation line character, the dislocation line within the identified slip plane is further approximated using cubic fitting. For that, the identified points $\bm{p}_{i,k}$ are projected onto the plane as
\begin{equation}
\bm{p}'_{i,k}=\bm{p}_{i,k}-\hat{\bm{n}}\left(\bm{p}_{i,k}\cdot\hat{\bm{n}}-\left|\bm{c}_j\right|\right).
\end{equation}
Parametrisation of the points using two perpendicular vectors $\hat{\bm{u}}$, and $\hat{\bm{v}}$ that  are contained in the slip plane $S_d$ results in
\begin{equation}
\begin{aligned}
\bm{p}'_{i,k}=s_i\hat{\bm{u}}+t_i\hat{\bm{v}}+\left|\bm{c}_j\right|\hat{\bm{n}} \\ \text{where} \quad \hat{\bm{u}}\perp\hat{\bm{v}}, \quad \hat{\bm{u}},\hat{\bm{v}}\in S_d.
\end{aligned}
\end{equation}
Least-square fitting of the scalar factors $\bm{s}=\left[s_1, ...,s_N\right]$ and $\bm{t}=\left[t_1, ...,t_N\right]$ using quadratic ansatz with coefficients $a$, $b$, $c$, and residual $r_i$ as 
\begin{equation}
s_i = ax_i^2+bx_i+c+r_i\text{, with }x_i=i/N\in[0,1]
\end{equation}
results in the coefficient vectors 
\begin{equation}
\bm{w}_u=\begin{bmatrix}a&b&c\end{bmatrix}\text{ and } \bm{w}_v \text{, respectively.}
\end{equation}
A point $\bm{d}_j\left(x\right)$ on the fitted dislocation line $d_n$ is hence evaluated as
\begin{equation}\label{eq:dislocation_line_analytic}
\bm{d}_j\left(x\right)=\begin{bmatrix}\hat{\bm{u}} & \hat{\bm{v}}\end{bmatrix}\begin{bmatrix}\bm{w}_u \\  \bm{w}_v\end{bmatrix}\begin{bmatrix}x^2\\x\\1\end{bmatrix}+ \left|\bm{c}_j\right|\hat{\bm{n}}
\end{equation}
To consolidate the reconstructions from different steps, say $\varepsilon_k$ and $\varepsilon_{k+1}$, it is assumed that dislocation migration only occurs within the slip plane. Hence, the centres of mass of a reconstructed dislocation lie on the same plane.\par
\begin{figure}[h]
    \centering
    \includegraphics[width=.75\linewidth]{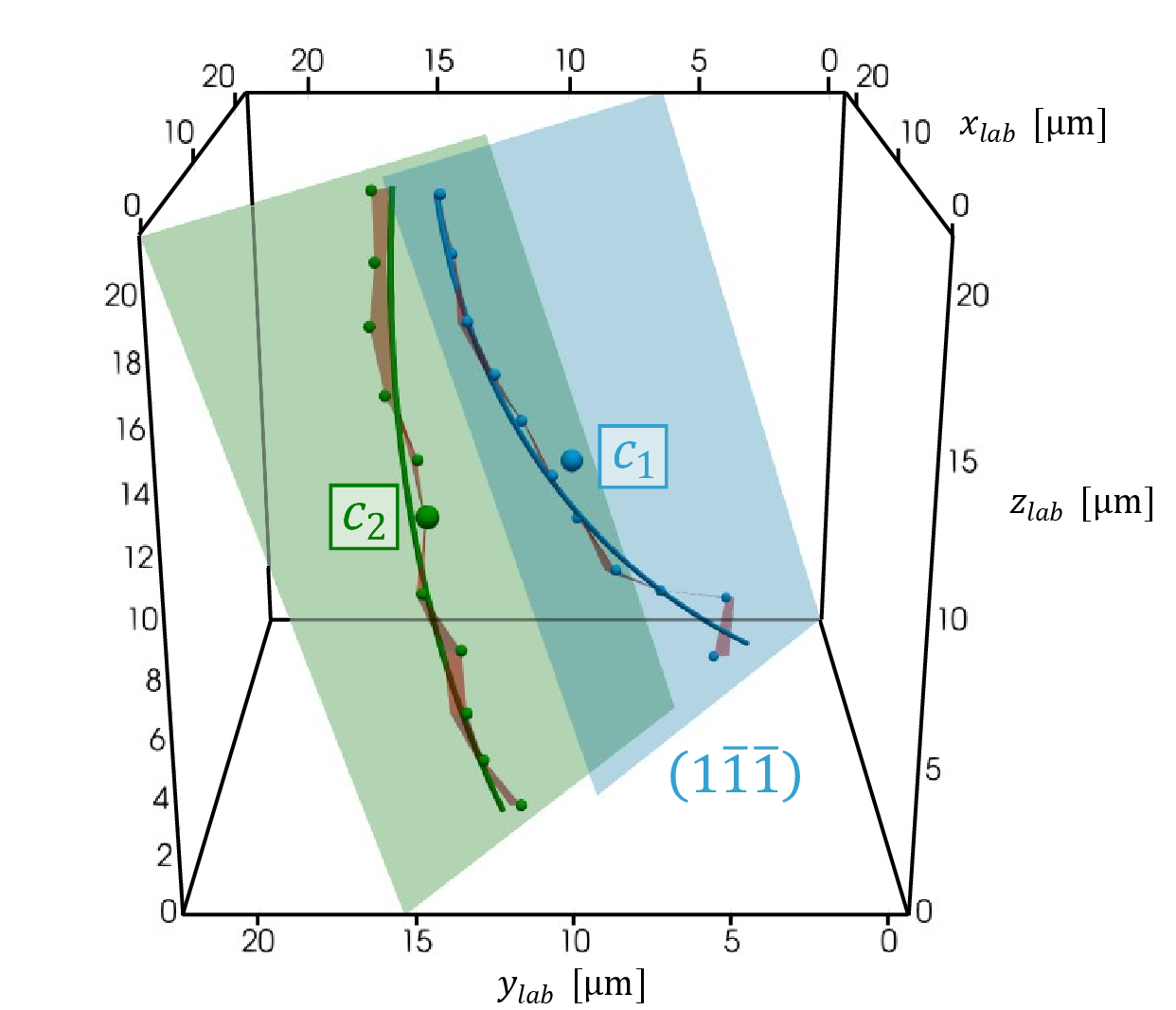}
    \caption{Reconstructed line fits of dislocation $10$ presented in Fig.~\ref{figSupp:2-intensityField}b) lying on a $\left(1\bar{1}\bar{1}\right)$ slip plane for steps $\varepsilon_4$ (blue) and $\varepsilon_5$ (green). The small dots indicate the dislocation positions $\bm{p}_{i,k}$ identified from the intensity field using Gaussian Blob Detection. The large spheres are their respective centres of mass $\bm{c}_j$. The red areas indicate the offsets between the identified points $\bm{p}_{i,k}$ and their projections $\bm{p}'_{i,k}$ onto the identified slip planes along the plane normal direction.}
    \label{figSupp:3-dislocationLinePlaneFit}
\end{figure}

An example result of the outlined reconstruction scheme for dislocation line 10, shown in Fig.~\ref{figSupp:2-intensityField}b, is presented for two consecutive steps $\varepsilon_4$ and $\varepsilon_5$ in Fig.~\ref{figSupp:3-dislocationLinePlaneFit}. To quantify the reconstruction method's accuracy, the mean squared error between the points and their fitted plane is computed. For dislocation 10, the error amounts to $\mathrm{MSE}=\SI{0.56}{\micro\metre}$.

\subsection{Interaction Force Calculation}\label{sec:appendix-force}
The force calculation is conducted on a per-segment basis. The segments are obtained through sampling of the analytic expressions of the dislocation lines (cf. eq.~\ref{eq:dislocation_line_analytic}). Each segment $s_i$ is defined by its start and end point, $\bm{p}_{i}^{\mathrm{start}}$ and $\bm{p}_{i}^{\mathrm{end}}$, and the dislocation line specific slip plane normal, $\bm{n}$ and Burgers vector $\bm{b}$. The force $\bm{F}_i$ on a segment is approximated by using the stress $\bm{\sigma}_i$ at its midpoint $\bm{p}_i^{\mathrm{mid}}=0.5(\bm{p}_{i}^\mathrm{start}+\bm{p}_{i}^\mathrm{end})$ as
\begin{equation}
    \frac{\bm{F}_i}{L_i}=\left(\bm{b_i}\cdot\bm{\sigma}_i\right)\times\hat{\bm{\xi}_i}.
\end{equation}
Here, $L_i$ and $\hat{\bm{\xi}_i}$ refer to the segment length and the normalised segment tangent, respectively, defined as
\begin{align}
    L_i&=\|\bm{p}_i^\mathrm{end}-\bm{p}_i^\mathrm{start}\| \\
    \hat{\bm{\xi}_i} &= L_i^{-1}(\bm{p}_i^\mathrm{end}-\bm{p}_i^\mathrm{start}).
\end{align}
The midpoint stress is obtained by superposing all long-range stress contributions from the other dislocation segments $\bm{\sigma}_{j\neq i}\left(\bm{p}_i^{\mathrm{mid}}\right)$ in the domain, and the applied load tensor $\bm{\sigma}_\mathrm{app}$ as
\begin{equation}
    \bm{\sigma}_i =\bm{\sigma}_\mathrm{app} +  \sum_{j\neq i}\bm{\sigma}_j\left(\bm{p}_i^{\mathrm{mid}}\right).
\end{equation}
The long-range elastic stress fields of the finite straight dislocation segments $\bm{\sigma}_{j\neq i}\left(\bm{p}_i^{\mathrm{mid}}\right)$ are evaluated using the coordinate-independent form presented in B.2. of \cite{Cai2006}.
The applied load tensor $\bm{\sigma}_\mathrm{app}$ represents the applied tensile load $\sigma_z$ along the sample's z-axis. Due to the tilt of the sample with respect to the reference laboratory frame of $\theta=\SI{-8.9}{\degree}$ around the reference frame y-axis, the applied load tensor is expressed as
\begin{equation}
    \bm{\sigma}_\mathrm{app} = \bm{\Theta}\bm{\sigma}_\mathrm{ext}\bm{\Theta}^{T}
\end{equation}
with
\begin{equation}
    \bm{\Theta} = 
\begin{bmatrix}
\cos\theta &  0 &\sin\theta \\
0 & 1 & 0 \\
-\sin\theta & 0 & \cos\theta 
\end{bmatrix}
\end{equation}
and 
\begin{equation}
    \bm{\sigma}_\mathrm{ext} = 
\begin{bmatrix}
0 & 0 & 0 \\
0 & 0 & 0 \\
0 & 0 & \sigma_z
\end{bmatrix}.
\end{equation}
Once the force on a segment $\bm{F}_i$ is obtained, the glide force on its slip plane is obtained through projection as
\begin{equation}
\bm{F}_i^\mathrm{glide}=\frac{\bm{F}_i\cdot\left(\hat{\bm{\xi}_i}\times\left[\bm{b}\times\hat{\bm{\xi}_i}\right]\right)}{\|\bm{b}\times\hat{\bm{\xi}_i}\|}
\end{equation}

\section{Supplementary Figures}\label{secA1}
\subsection{Pile-up Evolution}\label{sec:appendix-evolution}
\begin{figure}[h]
    \centering
    \includegraphics[width=1.0\linewidth]{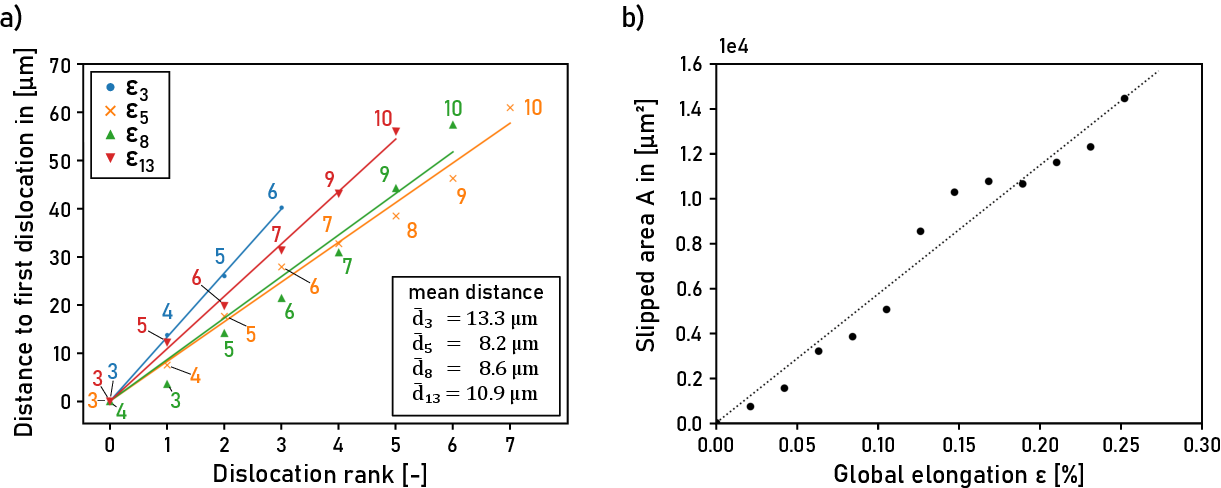}
    \caption{Statistical metrics of pile-up evolution. a) dislocation spacing as a function of rank in the pile-up for steps presented in Fig.~\ref{figSupp:5-evolution}. b) signed accumulated slip area of dislocation segments as a function of macroscopic applied elongation. The swept area per step is here defined as the total area enclosed by the reconstructed positions of each dislocation line across two consecutive steps. For each dislocation, the area is counted positively if the line migrates toward the pile-up head, and negatively if it moves toward the tail. The plotted values represent the cumulative swept area over the course of deformation. Global elongation is determined from the traverse displacement of the load frame. }
    \label{figSupp:4-structuralMetrics}
\end{figure}

\begin{figure}[h]
    \centering
    \includegraphics[width=1.0\linewidth]{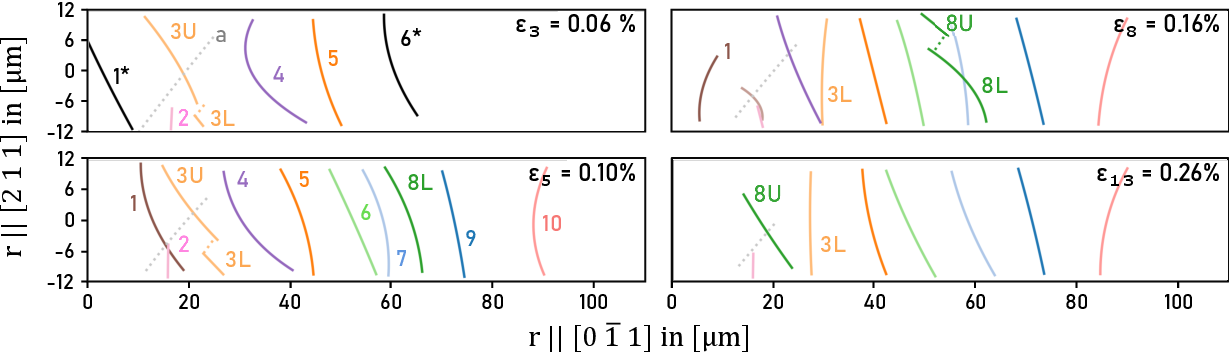}
    \caption{Dislocation configuration in pile-up viewed along the normal direction of the average habit plane (cf. red plane in Fig.~\ref{fig1:pile-upOverview}b)}
\label{figSupp:5-evolution}
\end{figure}

Labelled dislocation positions in the pile-up plane are presented for four steps in Fig.~\ref{figSupp:5-evolution}. The first projection shown in the figure marks the formation of the pile-up consisting of six dislocations at step $\varepsilon_3$ with a mean spacing of \SI{13.3}{\micro\metre} (cf. Fig.~\ref{figSupp:4-structuralMetrics}a). In the next shown step $\varepsilon_5$, the amount of dislocations increases to 10 at a substantially lower mean spacing of \SI{8.2}{\micro\metre}, indicating compression of the structure as new dislocations follow. As a result of the decreased mean spacing between the dislocation lines and increased interaction forces, dislocation \textit{8} is splitting up with the upper segment migrating to a parallel plane at a normal distance of \SI{2.4}{\micro\metre} from the pile-up through \text{cross-slip}. This separation precedes dislocation \textit{8} being dragged parallel to the pile-up towards the obstacle and being eventually pinned at dislocation \textit{a}. The gap is subsequently closed through relaxation of the structure to a mean spacing of \SI{10.9}{\micro\metre} as shown for step $\varepsilon_{13}$. Eventually, in step $\varepsilon_{15}$, the structure is dissolved.\par

\begin{figure}
    \centering
    \includegraphics[width=1\linewidth]{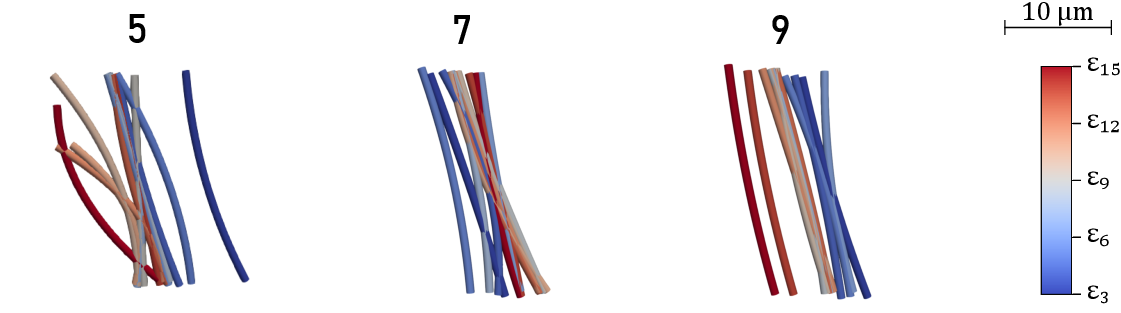}
    \caption{Overlay of the positions of dislocation lines 5, 7 and 9 for all steps showing non-linear motion.}
    \label{figSupp:6-positionOverlay}
\end{figure}

\begin{figure}
    \centering
    \includegraphics[width=.75\linewidth]{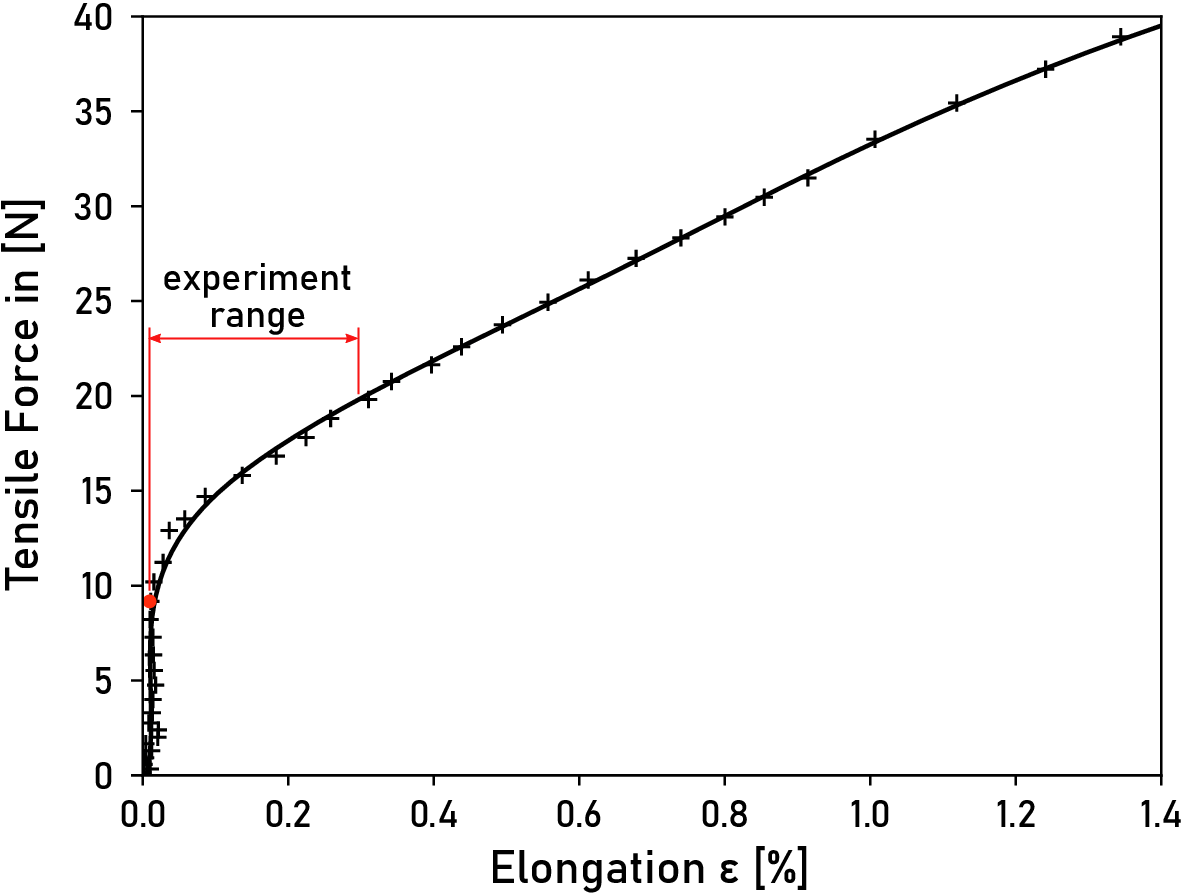}
    \caption{Force-elongation curve of a single crystal from the same batch with identical crystalline orientation. Elongations are determined by Digital Image Correlation (DIC). The range covered in this study is annotated for reference.}
    \label{figSupp:7-stress_strain_curve}
\end{figure}
\subsection{Dislocation Forces}\label{sec:app-forces}
\begin{figure}[h]
    \centering
    \includegraphics[width=1.0\linewidth]{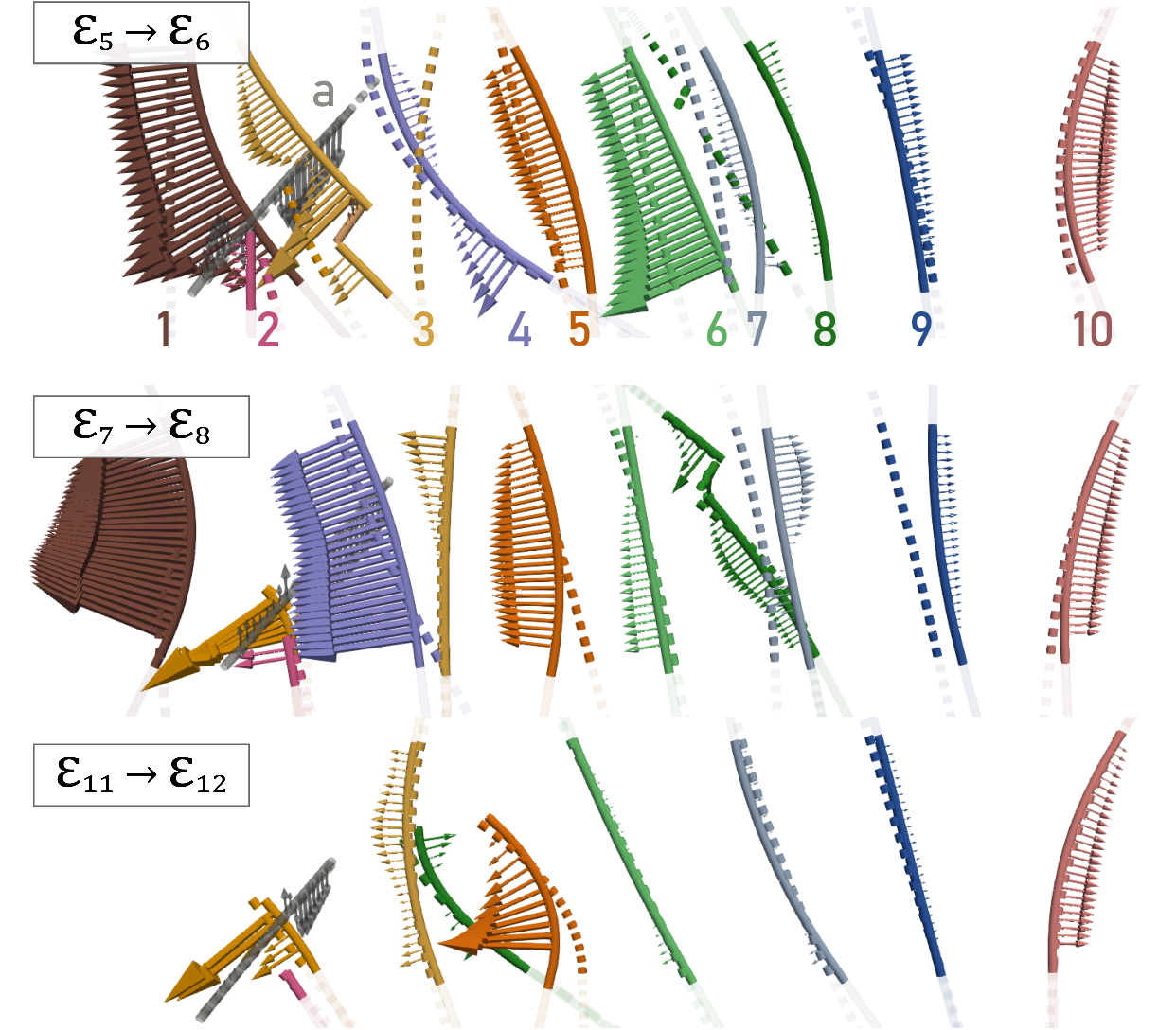}
    \caption{Result of the glide force calculation of the dislocation configuration in steps $\varepsilon_5$, $\varepsilon_7$ and $\varepsilon_{11}$. Dislocation lines are coloured as in Fig.~\ref{fig1:pile-upOverview}. Full lines represent the configuration used for the force calculation in step $\varepsilon_i$; dotted lines represent the subsequent step $\varepsilon_{i+1}$. The arrows are scaled by the magnitude of the glide force.}
\label{figSupp:8-mechanicalModelling}
\end{figure}

The forces on the dislocations based on the configurations in steps $\varepsilon_5$, $\varepsilon_7$, and $\varepsilon_{11}$ are shown in Fig.~\ref{figSupp:8-mechanicalModelling}. For all three shown steps, the calculated forces generally increase towards the obstacle, indicating the superposition of the applied load driving the dislocation towards the obstacle and the repulsive dislocation-dislocation interaction. Comparing the migration and the forces on a per-dislocation basis, however, reveals discrepancies between the model and observed motion, indicating a more complex stress landscape. For some dislocations, for example, for dislocations \textit{1}, \textit{5}, \textit{6} in step $\varepsilon_5$, even on a per-segment basis, the predictions of the model are in good agreement in terms of both direction and magnitude. Other predictions, for example, for dislocations \textit{10} and \textit{3}, are inaccurate at all steps. Changes in the force magnitude and dislocation migration distance of the same dislocation between steps (cf. dislocations 4 and 5) indicate a substantial influence of pinning mechanisms, most likely originating from effects outside the mapped volume.\par 
\renewcommand{\figurename}{Video}
\begin{figure}[h]
    \centering
    \includegraphics[width=0.5\linewidth]{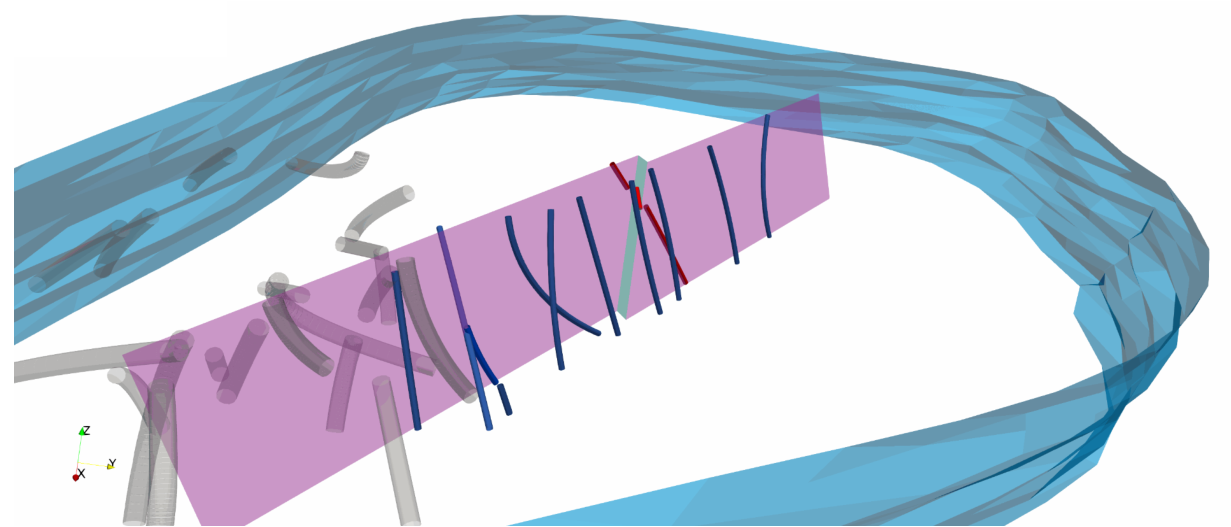}
    \caption{Snapshot from supplementary video illustrating the evolution of the dislocations in the pile‑up during the loading sequence, along with the slip planes of the cross‑slipping dislocation 8 (shown in red).}
\label{figSupp:9-Movie}
\end{figure}
\end{appendices}
\end{document}